\documentclass[conference,onecolumn]{IEEEtran}
\IEEEoverridecommandlockouts
\usepackage{cite}
\usepackage{amsmath,amssymb,amsfonts}
\usepackage{algorithmic}
\usepackage{graphicx}
\usepackage{textcomp}
\usepackage{xcolor}
\usepackage{wrapfig}
\def\BibTeX{{\rm B\kern-.05em{\sc i\kern-.025em b}\kern-.08em
    T\kern-.1667em\lower.7ex\hbox{E}\kern-.125emX}}
\begin{document}

\title{Optical Fiber Communication Systems Based on End-to-End Deep Learning\\
\thanks{The work was carried out under the EU Marie Sk{\l}odowska-Curie project COIN (No. 676448) and UK EPSRC programme grant TRANSNET (EP/R035342/1).}
}

\author{
  \IEEEauthorblockN{
    Boris Karanov\IEEEauthorrefmark{1}\IEEEauthorrefmark{3},
    Mathieu Chagnon\IEEEauthorrefmark{3},
    Vahid Aref\IEEEauthorrefmark{3},
    Domani\c{c} Lavery\IEEEauthorrefmark{1},
    Polina Bayvel\IEEEauthorrefmark{1},
    and Laurent Schmalen\IEEEauthorrefmark{2}
  }
  \vspace*{1ex}
  \IEEEauthorblockA{
    \IEEEauthorrefmark{1}Dept. Electronic \& Electrical Engineering, University College London, WC1E 7JE London, U.K. 
  }
  \IEEEauthorblockA{
    \IEEEauthorrefmark{2}Communications Engineering Lab, Karlsruhe Institute of Technology (KIT), 76131 Karlsruhe, Germany
  }
  \IEEEauthorblockA{
    \IEEEauthorrefmark{3}Nokia Bell Labs, 70435 Stuttgart, Germany
  }
}

\IEEEspecialpapernotice{(Invited Paper)}

\maketitle\vspace*{-1ex}

\begin{abstract}
We investigate end-to-end optimized optical transmission systems based on feedforward or bidirectional recurrent neural networks~(BRNN) and deep learning. In particular, we report the first experimental demonstration of a BRNN auto-encoder, highlighting the performance improvement achieved with recurrent processing for communication over dispersive nonlinear channels.
\end{abstract}\vspace*{-1ex}


\section{Introduction}
Deep learning techniques offer the opportunity to fundamentally reconsider the conventional communication system design by implementing (parts of) the transceiver using artificial neural networks (ANN). The parameters of these ANNs can be jointly tuned such that the end-to-end system performance is optimized, by interpreting the transceiver chain as an ANN-based deep \emph{auto-encoder}~\cite{O'Shea,Doerner}. Such fully learnable systems have attracted great interest in communication scenarios where the optimum transmitter-receiver pair is unknown or is infeasible to implement due to complexity constraints. An important example is the transmission over dispersive nonlinear channels such as the ubiquitously deployed optical fiber links, where the end-to-end deep learning has been recently introduced and experimentally verified~\cite{Karanov_1,Li,Jones}. In particular, auto-encoders were extensively studied for application in low-cost optical fiber communication systems based on intensity modulation and direct detection (IM/DD)~\cite{Karanov_1,Karanov_2}, a preferred technology in many data center, metro and access networks. The interplay of chromatic dispersion, introducing intersymbol interference (ISI), and nonlinear photodetection imposes severe performance limitations in the IM/DD links. In~\cite{Karanov_1}, the first experimental demonstration of an optical fiber auto-encoder showed that a system based on a feedforward ANN (FFNN) can outperform conventional pulse amplitude modulation (PAM) transmission with classical linear equalization. More recently, a transceiver tailored to the dispersive channel properties was proposed using a bidirectional recurrent ANN and sliding window sequence estimation (SBRNN)~\cite{Karanov_2}, extending the design of~\cite{Farsad} to auto-encoders. Simulation results showed that the SBRNN system has an improved
\begin{wrapfigure}[9]{l}{0.515\textwidth}\vspace*{-2.5ex}
\centering
\includegraphics[width=0.525\textwidth, keepaspectratio = true]{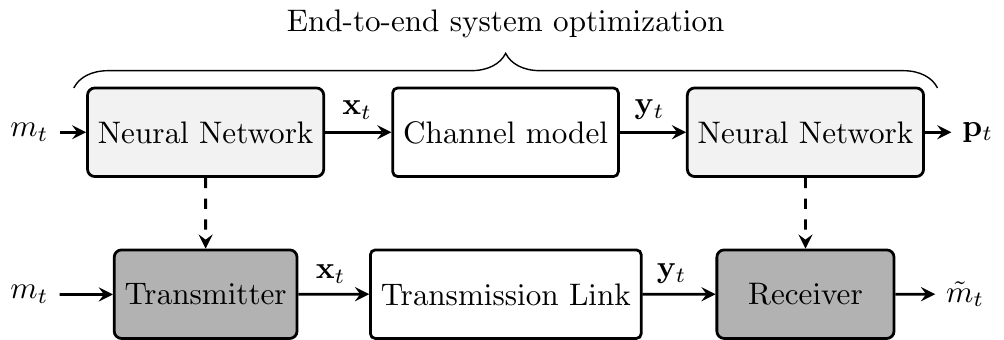}
\vspace*{-4ex}
\caption{Implementation of end-to-end system learning using a channel model.}
  \label{fig:end-to-end_schematic}
\end{wrapfigure} 
performance with lower complexity than PAM transmission with nonlinear equalizers~\cite{Karanov_2} or maximum likelihood sequence detection~\cite{Karanov_3}. In this paper, we review the FFNN and SBRNN designs and report results from the experimental demonstration of the SBRNN auto-encoder. Using the simple procedure of training the transceiver on a numerical channel model and applying it 'as is' to the experimental test-bed, a substantial reach increase is achieved over FFNN for a reference 42\,Gb/s transmission.

\begin{figure*}[t!]
\centering
\includegraphics[width=\textwidth, keepaspectratio=true]{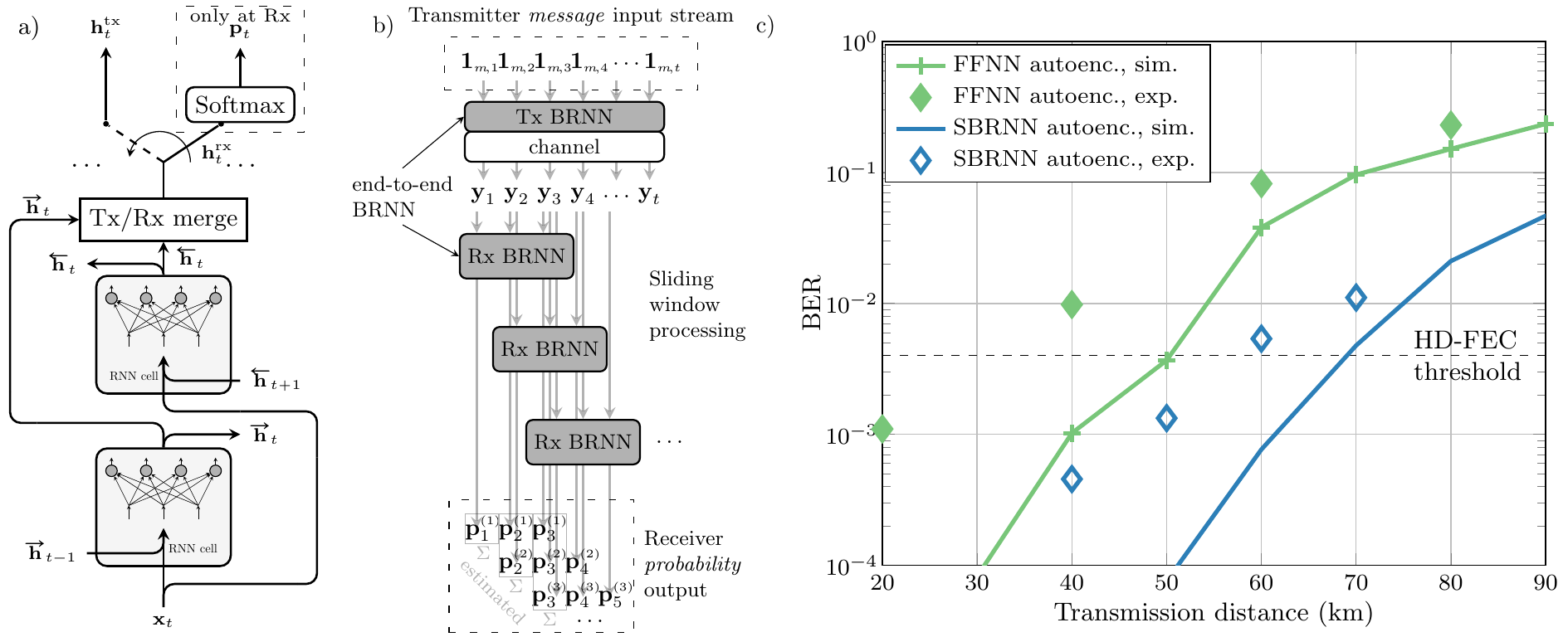}
\vspace*{-4ex}
\caption{\label{fig:Schematics_and_results}a) BRNN processing. b) Sliding window sequence estimation algorithm. c) BER versus transmission distance for FFNN and SBRNN auto-encoders.}\vspace*{-3.5ex}
\end{figure*}

\section{Optical Fiber Auto-encoder: Implementation and Performance}
Fully learnable communication systems are based on the idea of implementing the complete chain of transmitter, channel and receiver as an end-to-end deep artificial neural network~\cite{O'Shea,Doerner}. Using deep learning, the transmitter and receiver sections are trained jointly, such that we obtain a set of ANN parameters for which the end-to-end system performance is optimized for a specific metric, e.g. the symbol error rate. Figure~\ref{fig:end-to-end_schematic} shows the schematic of a practical implementation scheme for the auto-encoders. The transceiver is optimized in simulation using a numerical model of the actual channel and applied 'as is' to the transmission link. Such systems present a viable perspective for low-cost optical communications since their deployment does not require implementation of a training process in the final product. In this work, the modeling of the IM/DD link components is identical to~\cite[Sec.~III-B]{Karanov_1} and~\cite[Sec.~2.1]{Karanov_2}, while the experimental test-bed is as described in ~\cite[Sec.~V]{Karanov_1}. Previously~\cite{Karanov_1}, a block-based optical auto-encoder was experimentally demonstrated, each block representing an independently encoded and decoded message of a few data bits. This allowed a simple FFNN design, which, however, was inherently unable to compensate any ISI outside of the block. A more advanced system based on sequence processing by an end-to-end BRNN and sliding window estimation was proposed in~\cite{Karanov_2}, addressing the FFNN limitations. Figure~\ref{fig:Schematics_and_results}-a) and~-b) shows schematics of the BRNN and the sliding window algorithm, respectively, for which detailed descriptions can be found in~\cite{Karanov_2,Karanov_3}. The FFNN and BRNN outputs $\mathbf{h}_{t}^{\textnormal{FFNN}}$ and $\mathbf{h}_{t}^{\textnormal{BRNN}}$, respectively, can be expressed as (assuming a single FFNN layer for ease of exposition)
\begin{equation*}
\mathbf{h}_{t}^{\textnormal{FFNN}}=\alpha\left(\mathbf{W} \mathbf{x}_t+\mathbf{b}\right), \qquad \mathbf{h}_{t}^{\textnormal{BRNN}}=\textnormal{merge}_{\textnormal{tx/rx}}\left\{\alpha\left(\mathbf{W}_{\textnormal{fw}}\begin{pmatrix} \mathbf{x}_t^T & \overrightarrow{\mathbf{h}}_{t-1}^T\end{pmatrix}^T +\mathbf{b}_{\textnormal{fw}}\right);\alpha\left(\mathbf{W}_{\textnormal{bw}}\begin{pmatrix} \mathbf{x}_t^T & \overleftarrow{\mathbf{h}}_{t-1}^T\end{pmatrix}^T +\mathbf{b}_{\textnormal{bw}}\right)\right\},
\end{equation*}
where $\mathbf{x}_t$ is the neural network input at time $t$, $\alpha$ is an activation function, $\mathbf{W}$ and $\mathbf{b}$ denote weights and biases, and the \emph{merge} function combines the forward and backward processing of the BRNN, e.g. via concatenation or element-wise averaging. Note that in contrast to the FFNN, the BRNN handles the current input together with the preceding and succeeding output, enabling processing of data sequences. Figure~\ref{fig:Schematics_and_results}-c) compares the BER performance of the two auto-encoders at 42\,Gb/s as a function of distance in both simulation and experiment. Training of the transceiver was performed as described in~\cite[Sec. III-D]{Karanov_1} for FFNN and~\cite[Sec. 2.2]{Karanov_2} for SBRNN, aimed at minimizing the cross entropy between transmitter inputs and receiver outputs and thus improving the symbol error rate. We see that the SBRNN system significantly outperforms the simple FFNN, allowing transmission below the $6.7\%$ HD-FEC at distances beyond 50\,km in experiment. It is important to mention that the SBRNN performance can be further improved by a carefully chosen sliding window size~\cite[Sec. 3]{Karanov_2}. Moreover, we see that the transceivers, which are learned on a prior assumption of a specific channel model, achieve slightly deteriorated experimental performance as a result of any discrepancy between the model and the actual link as first reported in~\cite{Doerner}. Optimization of the receiver~\cite[Sec. VI-C]{Karanov_1} or the end-to-end fiber-optic system~\cite{Karanov_4} on collected experimental data can be employed to fine-tune the parameters of both the FFNN and SBRNN systems, trading-off additional implementation complexity for performance gains.

\section{Conclusions}
We report the results from the first experimental demonstration of an optical auto-encoder based on a recurrent neural network. In particular, we compare the performance of the SBRNN and the previous FFNN transceiver for the simple scenario where parameters are trained on a channel model and applied to the transmission link. In an excellent agreement with simulation, the SBRNN auto-encoder, owing to its processing of data sequences, allowed to increase system reach or enhance the data rate.

\end{document}